\documentstyle[aps,multicol]{revtex}
\begin{document}
\draft
\bibliographystyle{prsty}
\title{Micromagnetic simulations of thermally activated 
magnetization reversal of nanoscale magnets}
\author{Gregory Brown$^{1,2}$, 
        M. A. Novotny$^{1}$, 
    and Per Arne Rikvold$^{1,2}$}
\address{
$^1$Supercomputer Computations Research Institute,\\
$^2$Center for Materials Research and Technology,
and Department of Physics\\
Florida State University, Tallahassee, Florida 32306-4130
}

\date{\today}
\maketitle

\begin{abstract}

Numerical integration of a stochastic Landau-Lifshitz-Gilbert equation
is used to study dynamic processes in single-domain nanoscale magnets
at nonzero temperatures. Special attention is given to including
thermal fluctuations as a Langevin term, and the Fast Multipole Method
is used to calculate dipole-dipole interactions.  It is feasible to
simulate these dynamics on the nanosecond time scale for spatial
discretizations that involve on the order of $10^4$ nodes using a
desktop workstation. The nanoscale magnets considered here are single
pillars with large aspect ratio. Hysteresis-loop simulations are
employed to study the stable and metastable configurations of the
magnetization. Each pillar has magnetic end caps. In a time-dependent
field the magnetization of the pillars is observed to reverse via
nucleation, propagation, and coalescence of the end caps. In particular,
the end caps propagate into the magnet and meet near the middle. A
relatively long-lived defect is formed when end caps with opposite
vorticity meet. Fluctuations are more important in the reversal of the
magnetization for fields weaker than the zero-temperature coercive
field, where the reversal is thermally activated. In this case, the
process must be described by its statistical properties, such as the
distribution of switching times, averaged over a large number of
independent thermal histories.

\end{abstract}

\begin{multicols}{2}
The effect of temperature on the switching behavior of nanoscale
magnets can be quite strong when external fields are applied that are
just below the zero-temperature coercive threshold. Under these
conditions, thermal fluctuations can provide enough energy to take the
magnetization of the system over the barrier that prevents it from
aligning with the external field \cite{RIKV}. These issues are
important for understanding data integrity and high-speed switching in
single-domain magnetic applications.

Nanoscale magnets are modeled using the traditional
Landau-Lifshitz-Gilbert equation \cite{AHARONI},
\begin{equation}
\frac{ d {\bf M}_i}{dt} = - \frac{\gamma_0}{1+\alpha^2} {\bf M}_i
\times \left( {\bf H_i} - \frac{\alpha}{M_S}{\bf M}_i \times {\bf H}_i
\right)
\;,
\end{equation}
whereby the microscopic dipoles ${\bf M}_i$ precess under their
individual, locally observed applied fields ${\bf H}_i$.  The universal
gyromagnetic factor is $\gamma_0$=$1.76 \times 10^7$~Hz/Oe, while the
material parameters were selected to match those of bulk iron with a
saturation magnetization $M_S $=$ 1700~{\rm emu/cm^3}$, exchange length
$l_x $=$ 2.6~{\rm nm}$, and damping parameter $\alpha $=$ 0.1.$ For the
results considered here, these fields are composed of contributions
from a uniform field external to the system, the exchange interaction
with neighboring dipoles, and dipole-dipole interactions with all of
the other dipoles in the system.  Calculation of the latter is
dramatically accelerated by using a fast multipole algorithm
\cite{GREE87}. The numerical models are based on real nanomagnets that
have been fabricated recently using scanning microscopy techniques
\cite{WIRT98}.

Thermal effects are incorporated by adding a random contribution to
the local field of each spin, as first proposed by W. F. Brown
\cite{BROW63} nearly forty years ago. He calculated the
fluctuation-dissipation theorem for a Stratonovich-type \cite{KAMP}
noise by considering the Fokker-Planck formulation of the problem.
Numerically we implemented this noise using an It\^o-type \cite{KAMP}
random field. The difference between the Stratonovich and It\^o
paradigms does not affect the results since we normalize the dipoles
to a fixed length after each integration step.

The individual magnets, rectangular with dimensions $9~{\rm nm} \times
9~{\rm nm } \times 150~{\rm nm},$ were implemented using a
finite-difference approach.  The numerical results have been found to
be independent of discretization for the spatial discretization of
$\Delta x $=$ 1.5~{\rm nm}$ and integration step of $\Delta t $=$
50~{\rm fs}$ used here.

The large shape anisotropy of these magnets causes spontaneous
alignment of the magnetization, except at the top and bottom where
pole avoidance leads to the formation of end caps. As can be seen in
Fig.~1(a), when the external field is applied in the opposite
direction during a hysteresis measurement, the $z$-component of the
magnetization is reduced via growth of these end caps. Here the uniform
applied field points down, light shades indicate upward-pointing
magnetization, and dark shades indicate downward-pointing
magnetization. At zero temperature the end caps grow symmetrically;
when they meet at the middle of the pillar a relatively long-lived
defect is formed due to the opposite helicities of the two end caps.
These results are fully consistent with the $T=0$ simulations of
Ref.~\cite{YAN}. There the field was swept quasi-statically as opposed
to the truly dynamic sweeps considered here.

The corresponding hysteresis loop, with a period of $1$~ns, is shown
in Fig.~2. The defect is indicated by slow decay of the magnetization
around the time when the external field reverses. The effect of this
defect is also apparent for the longer-period hysteresis curves, for
which the defect disappears before the field reverses. The simulated
zero-temperature hysteresis loops are very reproducible, and no
differences are seen for subsequent periods. The inset shows example
variations that occur from thermal effects for the $1$~ns hysteresis
loop. The largest differences are seen in conjunction with the defect
that forms from the two end caps.

The simplest magnetization-reversal situation to study is one in which
the external field suddenly changes its orientation, and then remains
constant. In these simulations, the field is initially zero
and then is brought negative in $0.25$~ns with its
amplitude described by $1/4$ of a sine wave. In what follows we set
$t$=$0$ at the time when the field first reaches its maximum negative
value. When the final field is less than the coercive field, the
magnetization remains oriented upward until thermal fluctuations take
it over the associated free-energy barrier.  In long pillars the
end caps do not interact strongly, and the free energy of each can be
considered separately. The free energy as a function of end cap volume
has essentially three extrema: one local minimum corresponding to a
small end cap, one local maximum corresponding to an unstable volume
where the tendency for shrinkage is equal to the tendency for growth,
and the global minimum corresponding to a switched magnet (spanning
end cap).  Snapshots from the magnetization reversal of one magnet in
this situation at $H$=$-1850$~Oe and $20$~K are shown in
Fig.~1(b). Here the lower end cap undergoes a large fluctuation to take
it past the critical volume, after which it grows at an almost constant rate
to fill the entire magnet. (Again a long-lived defect forms when the
two end caps come into contact.)

The thermally-activated nature of the end cap growth leads to a
distribution in the switching times, defined as the time when
$M_z$=$0$. The probability of not switching, $P_{\rm not}(t)$, for $85$
switches at $H=-1800$~Oe and $20$~K is shown as the heavy, stepped
curve in Fig.~3. Under these conditions, the majority of simulated
switches occur between $0.5$ and $1.2$~ns after the field reversal
finishes. There are no switches before $0.4$~ns because it takes this
amount of time for a single supercritical end cap to grow to fill half
of the magnet. The traces of the average pillar magnetization density
as a function of time are shown for five different switches in the inset of
Fig.~3. There are essentially two slopes observed during the actual
switching process, corresponding to cases where one or both end caps
are growing.

The observation that the end caps decay essentially independently and
exponentially, with rate $\rho$, and that freely growing end caps
change the global magnetization (normalized to lie between $\pm 1$) at
a constant rate $v$, can be used to construct a simple model to
describe the distribution of switching times observed in individual
experiments. The resulting probability of not switching is
\end{multicols}
\begin{equation}
P_{\rm not}\left(t\right) = 
\left\{
\begin{array}{lr}
  1                                                         \qquad\qquad  & t<1/(2v) \\
  e^{-(2\rho t - \rho/v)} \left(1+2\rho t - \rho/v\right) \qquad\qquad  & 1/(2v)\le t< 1/v \\
  e^{-(2\rho t - \rho/v)} \left(1+          \rho/v\right)   \qquad\qquad  & {1/v \le t}
\end{array}
\right.
\;.
\end{equation}
\begin{multicols}{2}
\noindent
Taking $\rho$ and $v$ as parameters, a nonlinear fit of this
two-exponential decay theory to the first two moments of the data is
shown as the dashed line in Fig~3.  For comparison, a similar fitting
has also been performed for an error function form with two parameters
(corresponding to a Gaussian histogram of switching times), which is
shown as the dotted curve in Fig.~3. From the $85$ switches presented here there
is no clear advantage to either fitting function. Different
combinations of field and temperature should probe regions where the
fitting functions are not so similar.

In summary, Landau-Lifshitz-Gilbert dynamics have been simulated for
three-dimensional models of single-domain nanoscale magnets of large
aspect ratio. Hysteresis-loop and field-reversal simulations show
that magnetization reversal occurs through the nucleation, growth, and
coalescence of the end caps. For field-reversal simulations at nonzero
$T$ which require thermal fluctuations to complete the reversal, a
simple theory that considers the nucleation rate and growth velocity of
the end caps adequately describes the statistical distribution of the
switching times.

Supported by NSF grant No. DMR-9871455, NERSC, and by FSU/SCRI and
FSU/MARTECH.

\end{multicols}

\newpage
\begin{figure*}[t]
\null
\vskip 4in
\includegraphics{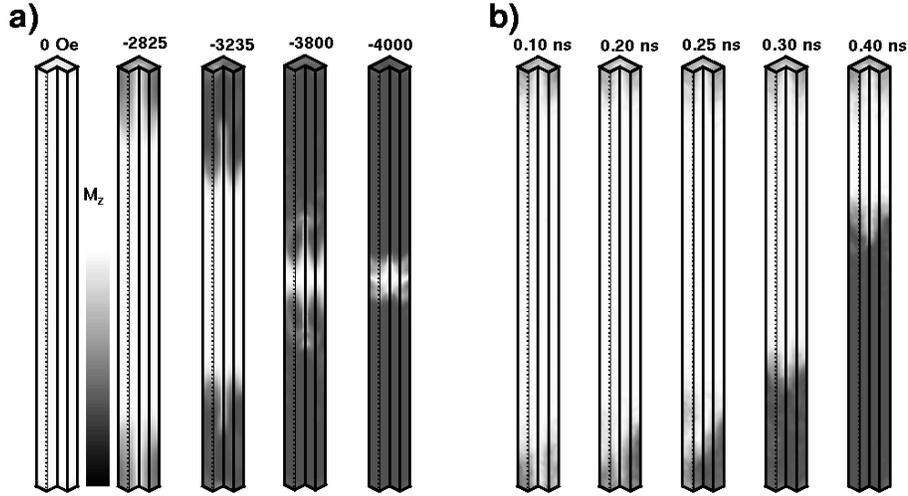}
\caption[]{Visualization of the $z$-component of the magnetization for
$9~{\rm nm} \times 9~{\rm nm } \times 150~{\rm nm}$ iron nanomagnets
(shown in $3/4$ cut-away view). Light shades indicate upward-pointing
magnetization, while dark shades indicate downward-pointing
magnetization. (a) Switching via symmetric growth of end caps for
$1$~ns period hysteresis loop at $0$~K. (b) Switching at $20$~K and
$H$=$1850$~Oe via thermal fluctuations of one end cap over the saddle
point, with subsequent growth to switch the magnetization. Movies are
available at {\tt http://www.scri.fsu.edu/$\sim$browgnrg/micromag/pillar.html}}
\end{figure*}

\begin{multicols}{2}

~
\begin{figure}[b]
\null
\vskip 2.15in
\includegraphics{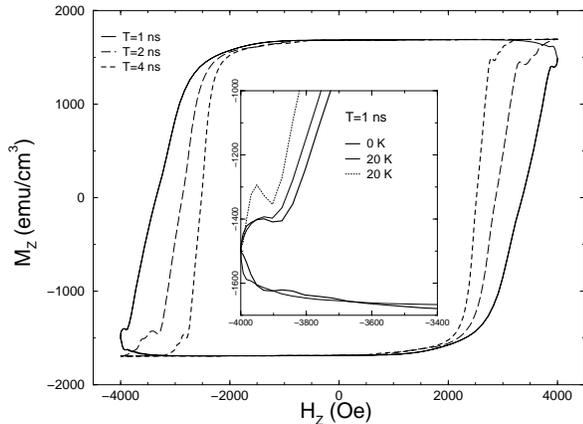}
\caption[]{Simulated hysteresis loops at $0$ K for applied fields
oscillating with periods of $1$, $2$, and $4$~ns. The inset shows
variations caused by temperature for two loops simulated at $20 K.$
The feature that occurs at nearly saturated magnetizations is due to a
defect that forms when the end caps come in contact.}
\end{figure}

\vskip 0.1in

~
\begin{figure}[b]
\null
\vskip 2.15in
\includegraphics{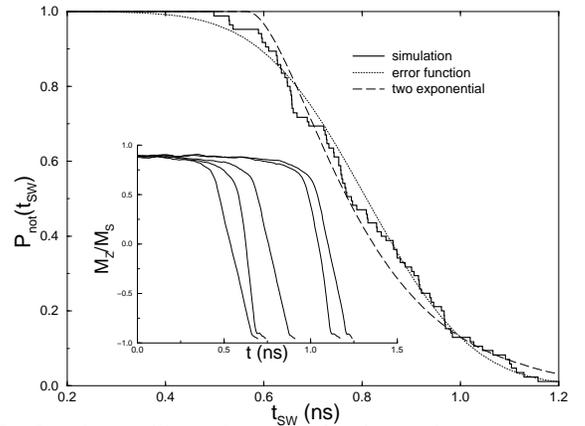}
\caption[]{Probability of not switching, $P_{\rm not}$,
for nanomagnets experiencing external fields just below the
coercive value. The solid line is simulation data from 85 switches at
$H$=$-1800$~Oe and $20$~K. Two theoretical forms [the dotted line is
an error function corresponding to a Gaussian histogram of switching
times, and the dashed line corresponds to the two-exponential decay
theory described in the text, Eq.~(2)] are compared after nonlinear fitting
with two parameters. The inset shows 5 examples of the average
$z$-component of the magnetization during different switching events.}
\end{figure}

\vskip 0.1in

\end{multicols}

\end{document}